\documentclass[aps,pra,longbibliography,numerical,reprint,onecolumn,a4paper,10pt,notitlepage,nofootinbib]{revtex4-1}

\usepackage[english]{babel}
\usepackage[utf8]{inputenc} 
\usepackage[T1]{fontenc}
\usepackage{amsmath} 
\usepackage{graphicx}
\usepackage{epstopdf}    

\usepackage[caption=false]{subfig}

\usepackage[inline]{enumitem} 

\usepackage[top=3.5cm,bottom=3.5cm,inner=2.5cm,outer=2.5cm,headsep=1.25cm,headheight=1cm,footskip=1.5cm,nomarginpar]{geometry}

\RequirePackage{parskip}
\usepackage[hyperindex=true,bookmarks=true]{hyperref} 
\hypersetup{
    colorlinks=true,
    linkcolor=blue,
    citecolor=blue,
    urlcolor=blue,
}
\usepackage[all]{hypcap}

\begin{document}

\title{\Large Coupled dynamics of node and link states in complex networks: \\ A model for language competition}

\author{Adri\'{a}n Carro}
\email{adrian.carro@ifisc.uib-csic.es}

\author{Ra\'{u}l Toral}

\author{Maxi {San Miguel}}

\affiliation{IFISC (CSIC-UIB), Instituto de F\'isica Interdisciplinar y Sistemas Complejos,
Campus Universitat de les Illes Balears, E-07122, Palma de Mallorca, Spain}

\begin{abstract}

Inspired by language competition processes, we present a model of coupled evolution of node and link states. In particular, we focus on the interplay between the use of a language and the preference or attitude of the speakers towards it, which we model, respectively, as a property of the interactions between speakers (a link state) and as a property of the speakers themselves (a node state). Furthermore, we restrict our attention to the case of two socially equivalent languages and to socially inspired network topologies based on a mechanism of triadic closure. As opposed to most of the previous literature, where language extinction is an inevitable outcome of the dynamics, we find a broad range of possible asymptotic configurations, which we classify as: frozen extinction states, frozen coexistence states, and dynamically trapped coexistence states. Moreover, metastable coexistence states with very long survival times and displaying a non-trivial dynamics are found to be abundant. Interestingly, a system size scaling analysis shows, on the one hand, that the probability of language extinction vanishes exponentially for increasing system sizes and, on the other hand, that the time scale of survival of the non-trivial dynamical metastable states increases linearly with the size of the system. Thus, non-trivial dynamical coexistence is the only possible outcome for large enough systems. Finally, we show how this coexistence is characterized by one of the languages becoming clearly predominant while the other one becomes increasingly confined to ``ghetto-like'' structures: small groups of bilingual speakers arranged in triangles, with a strong preference for the minority language, and using it for their intra-group interactions while they switch to the predominant language for communications with the rest of the population.

\end{abstract}


\maketitle

\section{Introduction}
\label{C6_s:introduction}


Collective social phenomena have traditionally been studied with models based on node states \cite{Castellano2009}. While the analysis of dynamics based on link states has received increasing attention in the last decade \cite{Antal2005,Antal2006,Radicchi2007,Szell2010,Leskovec2010A,Leskovec2010B,Marvel2011,Traag2009,Evans2009,Evans2010,Ahn2010,Liu2012,Nepusz2012,FernandezGracia2012,Carro2014}, no attempt has so far been made to combine the two approaches. We address this lack by studying a model where both node and link states are taken into account and characterized by distinct but coupled dynamics. We focus on language competition, a process where both the characteristics of the speakers and the properties of their interactions play a relevant role.


The study of language competition in processes of language contact addresses the dynamics of language use in multilingual social systems due to social interactions. A main goal of the theoretical modeling of these processes is to distinguish between the mechanisms and conditions that lead to the coexistence of different languages and those leading to the extinction of all but one of them. The focus of the field is on language shift in terms of users, rather than changes in the language itself (for an evolutionary approach to the latter see \cite{Nowak1999,Nowak2002}). As a consequence, language is conceptualized as a discrete property of the speakers \cite{Castello2013}. In recent years, a number of contributions lying outside the realm of traditional sociolinguistics have addressed the problem of language competition from alternative perspectives, using tools and methods from statistical physics, nonlinear dynamics, and complex systems science \cite{Stauffer2006,Schulze2006,Loreto2007,Schulze2008,Baronchelli2012,Patriarca2012,Mufwene2016}. Much of this research stems from the seminal work by Abrams and Strogatz \cite{Abrams2003} about the dynamics of endangered languages. The model proposed by Abrams and Strogatz considers a binary-state society, where individuals can be either speakers of language $A$ or speakers of language $B$. Thus, by definition, this model can only account for societal bilingualism, i.e., for the coexistence of two different monolingual groups \cite{Appel2006}. Subsequent studies, on the contrary, have considered generalizations of the original model accounting for the existence of bilingual individuals \cite{Wang2005,Mira2005,Castello2006,Castello2007,Minett2008,Toivonen2009,Vazquez2010,Heinsalu2014}\footnote{Interestingly, an alternative modeling approach developed by Baggs and Freedman \cite{Baggs1990,Baggs1993} had already addressed the problem of individual bilingualism from a macroscopic, mean-field perspective more than a decade before the introduction of the Abrams-Strogatz model.}. In particular, the AB-model was proposed by Castell\'{o} \emph{et al} \cite{Castello2006} based on the works of Wang and Minett \cite{Wang2005}. It develops a modification of the original, binary-state Abrams-Strogatz model to account for the case of two non-excluding options by introducing a third, intermediate state ($AB$) to represent bilingual speakers.


An alternative, more natural way of accounting for bilingual speakers is to consider language as a property of the interactions between individuals \cite{FernandezGracia2012,Carro2014}. In fact, while traditionally conceptualized as a property of the speaker, the use of a language as a means of communication can be more clearly described as a feature of the relationship between two speakers \mbox{---a} link \mbox{state---} than as an attribute of the speakers themselves \mbox{---a} node \mbox{state---.} In this manner, bilingualism is not anymore an ad-hoc intermediate state, but the natural consequence of individuals using different languages in different interactions. Furthermore, this approach allows for a more nuanced understanding of bilingualism: speakers are not only characterized by being bilinguals or not, but by a certain degree of bilingualism, depending on the share of interactions in each language.


The study of models and dynamics based on link states has received increasing attention from areas of research such as social balance theory \cite{Heider1946,Antal2005,Antal2006,Radicchi2007,Szell2010,Leskovec2010A,Leskovec2010B,Marvel2011}, community detection \cite{Traag2009,Evans2009,Evans2010,Ahn2010,Liu2012}, and network controllability \cite{Nepusz2012}. While these models set a precedent in the use of a link-state perspective, they are not suitable for modeling the dynamics of competing languages: the two link states considered by social balance theory are not equivalent, friendship and enmity playing rather different roles; no dynamics of the link states has been studied in the context of community detection; and only continuous link states have been considered in network controllability problems. More relevant in the context of language competition, two recent contributions have addressed the study of link-state dynamics with binary, equivalent states. On the one hand, Fern\'{a}ndez-Gracia \emph{et al} implemented a majority rule for link states \cite{FernandezGracia2012}, finding a broad distribution of non-trivial asymptotic configurations characterized by the coexistence of both languages, including both frozen and dynamically trapped configurations. Interestingly, these non-trivial asymptotic configurations are found to be significantly more likely than under the traditional majority rule for node states in the same topologies \cite{Haggstrom2002,Castellano2005,Castellano2006,Baek2012}. On the other hand, Carro \emph{et al}~\cite{Carro2014} developed a coevolution model that couples the aforementioned majority rule dynamics of link states with the evolution of the network topology. Depending on the ratio between the probabilities of these two processes, the system is found to evolve towards different absorbing configurations: either a one-component network with all links in the same state \mbox{---extinction} of one of the \mbox{languages---} or a network fragmented in two components with opposite states \mbox{---survival} of both languages in completely segregated \mbox{communities---.}


We focus here on the fact that, while the use of a language can be clearly described as a property of the interactions between speakers \mbox{---link} \mbox{states---,} there are certain features intrinsic to these speakers \mbox{---node} \mbox{states---} which have a relevant influence on the language they choose for their communications. In particular, the attitude of a speaker towards a given language \mbox{---which} determines her willingness to use \mbox{it---} is affected by individual attributes such as her level of competence in that language, her degree of cultural attachment and affinity with the social group using that language, and the strength of her sense of identity or belonging to that group. For simplicity, we consider that all individual properties affecting language choice can be subsumed under the concept of ``preference''. At the same time, the evolution of the speakers' individual preferences is, in turn, affected by the languages used in their respective social neighborhoods. In this manner, the problem of language competition can be studied from the point of view of the intrinsically coupled evolution of language use and language preference. Ultimately, this change of perspective can be regarded as a shift from a paradigm in which language is considered only as a cultural or identity trait to one in which the tight entanglement with its role as a means of communication is also taken into account.


In order to address this intertwined dynamics of language use \mbox{---as} a means of \mbox{communication---} and language preference \mbox{---as} an attitude towards \mbox{it---,} we propose here a model of coupled evolution of node and link states. In particular, the use of two socially equivalent languages is represented by a binary-state variable associated to the links. In addition, nodes are endowed with a discrete real variable representing their level of preference for one or the other language. The dynamics of link states results from the interplay between, on the one hand, the tendency of speakers to reduce the cognitive effort or cost associated with switching between several languages \cite{Meuter1999,Jackson2001,Abutalebi2007,MoritzGasser2009} and, on the other hand, their tendency to use their internally preferred language. Regarding the dynamics of node states, we assume that speakers update their preference towards the language most commonly used in their respective social neighborhoods, i.e., the one most frequently used by their neighbors to communicate between themselves. In this way, node states influence the evolution of link states and vice versa.



In the above described mechanism of evolution of node states we implicitly assume that triangles \mbox{---sets} of three nodes connected with each \mbox{other---} represent actual group relationships, in which each speaker is aware of the interaction between the other two. Thus, we are led to focus on network topologies where triangles are abundant, a topic that has received a great deal of attention throughout the last decade \cite{Serrano2005,Newman2009,Bianconi2014}. Interestingly, it has been recently shown that real social networks are characterized by large clustering coefficients and, therefore, contain a large proportion of triangles \cite{Newman2003,Dorogovtsev2003,Newman2010,Foster2011,Colomer2013}. Triadic closure \cite{Rapoport1953} \mbox{---the} principle that individuals tend to make new acquaintances among friends of \mbox{friends---} has been found to be a successful mechanism in reproducing these structural properties. While a number of different implementations of the triadic closure mechanism have been studied in different contexts \cite{Holme2002,Davidsen2002,Sole2002,Vazquez2003,Boguna2004b,Krapivsky2005,Jackson2007}, we focus here on a socially inspired network generation algorithm proposed by Klimek and Thurner \cite{Klimek2013}, whose results have been validated with data from a well-known massive multiplayer online game \cite{Szell2010,Szell2010B,Szell2012,Klimek2016}. Note, nonetheless, that in order to have a well-defined evolution of speakers' preferences, we introduce a small modification to the algorithm so as to ensure that every node belongs to, at least, one triangle. Finally, note that we focus here on fixed topologies (for studies of coevolving networks see \cite{Zimmermann2001,Zimmermann2004,Eguiluz2005,Vazquez2008B,Jiang2011,Wang2014}).


A broad range of possible asymptotic configurations is found, which can be divided into three categories: frozen extinction of one of the languages, frozen coexistence of both languages, and dynamically trapped coexistence of both languages. Furthermore, metastable states with non-trivial dynamics and very long survival times are frequently found. The situations of coexistence (frozen, dynamically trapped, and metastable) are characterized by one of the languages becoming a minority but persisting in the form of ``ghetto-like'' structures, where predominantly bilingual speakers use it for the interactions among themselves \mbox{---mostly} \mbox{triangular---} but switch to the majority language for communications with the rest of the system \mbox{---generally} \mbox{non-triangular---.} A system size scaling shows, on the one hand, that the probability of extinction of one of the languages vanishes exponentially for increasing system sizes and, on the other hand, that the time scale of survival of the non-trivial dynamical metastable states increases linearly with the size of the system. Thus, non-trivial dynamical coexistence is the only possible outcome for large enough systems.


A detailed presentation of the model is given in Section~\ref{C6_s:the_model}, with a particular emphasis on the coupling between the dynamics of link states and the dynamics of node states. The structural constraints imposed by the definition of the model are also described in this section, as well as the particularities of the networks used for the numerical simulations. The different asymptotic configurations of the model, as well as their respective probabilities, are presented in Section~\ref{C6_s:transient_dynamics_and_asymptotic_configurations}, while in Section~\ref{C6_s:time_scales_of_extinction_and_metastable_coexistence} we study two different time scales characterizing the transient dynamics of the model before reaching these asymptotic configurations. In Section~\ref{C6_s:Use_of_the_minority_language} we investigate the role of bilingual speakers in sustaining the coexistence of both languages. A detailed comparison with the previously proposed $AB$-model, which also takes into account the existence of bilingual speakers, is addressed in Section~\ref{C6_s:Comparison_with_the_AB-model}. Finally, some conclusions are drawn in Section~\ref{C6_s:Conclusions}.


\section{The model}
\label{C6_s:the_model}

Inspired by the aforementioned language competition processes, we consider a population of $N$ speakers and the linguistic interactions between them \mbox{---any} social interaction mediated through \mbox{language---,} represented, respectively, by the nodes and the links of a network. We focus on the competition between two socially equivalent languages, that we label as $A$ and $B$. On the one hand, each speaker $i$, with $k_i$ neighbors in the network of interactions, is characterized by a certain preference $x_i$ for language $A$ (node state), being $(1-x_i)$ its preference for language $B$. In particular, we model the preference $x_i$ as a discrete variable taking values \mbox{$x_i \in \{ 0, 1/k_i, 2/k_i, \ldots , 1 \}$}, where $x_i=1$ indicates an absolute or extreme preference for language $A$ and $x_i=0$ an absolute or extreme preference for language $B$. On the other hand, each interaction between speakers can take place in one of the two possible languages, being thus each link $i$--$j$ characterized by a binary variable $S_{ij}$ (link state) taking the value $S_{ij} = 1$ if the language spoken is $A$ and $S_{ij} = 0$ if language $B$ is used.

Finally, the states of nodes and links evolve asynchronously, i.e., a single node or link is updated at each time step: with probability $p$ a randomly chosen node is updated, and with the complementary probability $(1-p)$ a randomly chosen link is updated. Therefore, the probability $p$ sets the relationship between the time scale of evolution of the speakers' preferences and the time scale at which the language used in conversations changes. Note that time is measured in the usual Monte Carlo steps, with $N$ updating events per unit time \mbox{---whether} node or link \mbox{updates---.} While the parameter $p$ does have an effect on how fast the system reaches its asymptotic behavior, the main features of this asymptotic regime seem to be unaffected by it \cite{Carro2016PHD}. Thus, we focus here on the particular case of equal probability of node and link updates, i.e., $p=0.5$.


\subsubsection*{Evolution of link states}

\begin{figure}[ht]
    \centering
    \includegraphics[width=8.0cm, height=!]{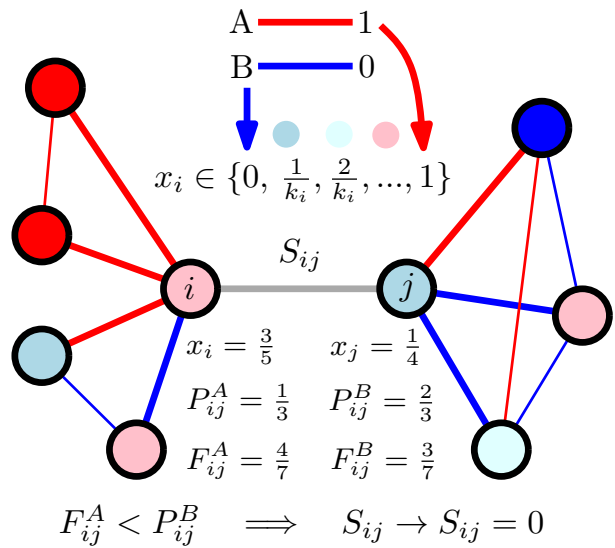}
    \caption{Schematic illustration of the evolution of link states. The use of the two competing languages $A$ and $B$ is represented, respectively, by red links and blue links, while the preferences of the speakers are represented by node colors ranging from red to blue through white. The interaction being updated is represented by a gray link. Only the links marked as bold are relevant for the particular link update illustrated here.}
    \label{C6_f:linkEvolution}
\end{figure}

The dynamics of link states \mbox{---the} language used in the interactions between \mbox{speakers---} results from the interplay between two mechanisms. On the one hand, we assume that there is a cognitive effort or cost associated with the use of several languages \cite{Meuter1999,Jackson2001}, which leads speakers to try to use the same language in all their conversations. As a consequence, the interaction between two given speakers tends to take place in the language most often used by both of them in their conversations with other speakers. In particular, we can define for each link $i$--$j$ the \emph{majority pressure} for language $A$ as the fraction of the total number $(k_i - 1) + (k_j - 1)$ of interactions with other speakers in which language $A$ is used,
\begin{equation}
    F_{ij}^A = \frac{k_i^A + k_j^A - 2S_{ij}}{k_i + k_j - 2} \, ,
    \label{C6_e:majorityPressure}
\end{equation}
where $k_i^A$ stands for the number of interactions in which speaker $i$ uses language $A$, and $k_i$ for its total number of interactions. On the other hand, speakers tend to use their internally preferred language: the higher their preference for a given language, the more willing they are to enforce its use in their conversations with other speakers. Combining the preferences of both participants in each interaction $i$--$j$, we can define the \emph{link preference} for language $A$ as
\begin{equation}
    P_{ij}^A =
    \begin{cases}
        \displaystyle\frac{x_i x_j}{x_i x_j + (1-x_i)(1-x_j)}\, , \quad & \text{if} \quad x_i x_j + (1-x_i)(1-x_j) > 0\, ,\\[25pt]
        \displaystyle\frac{1}{2}\, , & \text{otherwise}\, ,
    \end{cases}
    \label{C6_e:linkPreference}
\end{equation}
where the second case ensures that $P_{ij}$ is well-defined when there is a tie between two speakers with extreme preferences for different languages (\mbox{$x_i=0$} and $x_j=1$, or $x_i=1$ and $x_j=0$). This expression for $P_{ij}^A$ takes into account the preferences of both nodes, $x_i$ and $x_j$, in such a way that it yields the value $1$ if both nodes have complete preference for language $A$, $x_i = x_j = 1$, and it yields $0$ if both nodes have null preference for it, $x_i = x_j = 0$. If one of the nodes is neutral with respect to language $A$, say $x_i = 1/2$, then the link preference is equal to the other node's preference, $x_j$. Finally, the definition is such that it satisfies the requirement $P_{ij}^A (x_i, x_j) = 1 - P_{ij}^A (1 - x_i, 1 - x_j)$ reflecting the symmetry between the two languages. A schematic example of the calculation of these two quantities is shown in Fig.~\ref{C6_f:linkEvolution}. Note that the majority pressure and link preference for language $B$ are, respectively, $F_{ij}^B = 1 - F_{ij}^A$ and $P_{ij}^B = 1 - P_{ij}^A$.

When a link $i$--$j$ is picked for updating, its new state is chosen according to the following rules: 
\begin{enumerate*}[label=(\emph{\roman*})]
    \item if the sum of the majority pressure and the link preference for language $A$ is larger than the corresponding sum for language $B$, then language $A$ is chosen;
    \item if, on the contrary, the sum is larger for language $B$ than it is for language $A$, then language $B$ is chosen;
    \item if there is a tie between both languages, then one of them is chosen at random.
\end{enumerate*}
Given the aforementioned symmetry between both languages, these rules can be formally written as
\begin{equation}
    S_{ij} =
    \begin{cases}
        1\; (\text{language } A) \, , \quad & \text{if} \quad F_{ij}^A > P_{ij}^B \, ,\\[5pt]
        0\; (\text{language } B) \, , \quad & \text{if} \quad F_{ij}^A < P_{ij}^B \, ,\\[5pt]
        0 \text{ or } 1 \text{ randomly} \, , \quad & \text{if} \quad F_{ij}^A = P_{ij}^B \, ,
    \end{cases}
    \label{C6_e:linkEvolution}
\end{equation}
(see Fig.~\ref{C6_f:linkEvolution} for a schematic example of a link update). Note that, if the preferences of all the speakers are fixed as $1/2$, then all link preferences are also fixed as $1/2$ and we recover the majority rule for link states studied by Fern\'{a}ndez-Gracia \emph{et al} \cite{FernandezGracia2012}: the state of a link is updated to the state of the majority of its neighboring links. With freely evolving preferences of the speakers, on the contrary, the threshold for a state to be considered a majority is not anymore universal and fixed at $1/2$, but becomes local and dynamic: the fraction of neighbors in state $B$ needs to be larger than $P_{ij}^A$ for link $i$--$j$ to change its state to $B$, while the fraction of neighbors in state $A$ needs to be larger than $P_{ij}^B = (1 - P_{ij}^A)$ for it to change its state to $A$. Finally, note that speakers with extreme preferences ($x_i = 0$ or $x_i = 1$) impose their preferred language in all their conversations, except when they are faced by a speaker with an extreme preference for the other language, when there is an equilibrium between them and the language for their interaction is chosen at random. Thus, we are implicitly assuming that all speakers are able to use both languages, in line with the models previously proposed in the literature.


\subsubsection*{Evolution of node states}

\begin{figure}[ht]
    \centering
    \includegraphics[width=9.3cm, height=!]{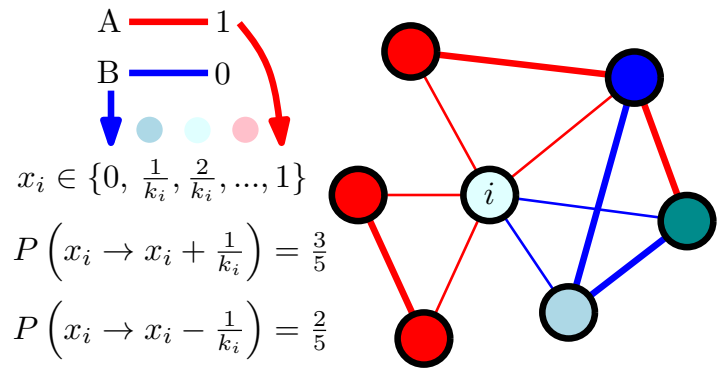}
    \caption{Schematic illustration of the evolution of node states. The use of the two competing languages $A$ and $B$ is represented, respectively, by red links and blue links, while the preferences of the speakers are represented by node colors ranging from red to blue through white. Only the links marked as bold are relevant for the particular node update illustrated here.}
    \label{C6_f:nodeEvolution}
\end{figure}

Regarding the dynamics of node states, we assume that speakers update their preferences according to the language that they observe their neighbors use between them \mbox{---obviously,} only those who are also neighbors of each \mbox{other---.} Thus, we implicitly assume that triangles represent actual group relationships, in which each speaker is aware of the interaction between the other two (see \cite{Serrour2011} for a study on the relationship between communities and triangles). In these terms, the more often the participants of the closer social group of a speaker \mbox{---her} triangular \mbox{relationships---} use a given language to communicate between themselves, the more likely it is that the speaker will update her preference towards that language. In particular, when a node $i$ is picked for updating, its state $x_i$ evolves according to the following probabilities
\begin{equation}
\begin{aligned}
    P\left(x_i \to x_i + \frac{1}{k_i}\right) &=
    \begin{cases}
        \displaystyle\frac{T_i^A}{T_i} \, , \quad & \text{if} \quad x_i \neq 1 \, ,\\[10pt]
        0 \, , \quad & \text{otherwise} \, ,
    \end{cases}\\[12pt]
    P\left(x_i \to x_i - \frac{1}{k_i}\right) &=
    \begin{cases}
        \displaystyle\left(1-\frac{T_i^A}{T_i}\right) \, , \quad & \text{if} \quad x_i \neq 0 \, ,\\[10pt]
        0 \, , \quad & \text{otherwise} \, ,
    \end{cases}
    \label{C6_e:nodeEvolution}
\end{aligned}
\end{equation}
where $T_i$ is the total number of links between neighbors of node $i$ and $T_i^A$ is the number of those links in state $1$, i.e., those in which language $A$ is used (see Fig.~\ref{C6_f:nodeEvolution} for a schematic example of a node update). Note that this evolution is equivalent to a one-dimensional random walk in the discrete-state space \mbox{$x_i \in \{ 0, 1/k_i, 2/k_i, \ldots , 1 \}$} with a bias towards $0$ or $1$ given by the probabilities in Eq.~\eqref{C6_e:nodeEvolution}. The fact that the modification of the preference ($\Delta x = 1/k_i$) is larger in nodes with fewer links can be motivated by noting that they tend to have fewer triangles and, therefore, each of them has a stronger influence on the node.

\subsubsection*{Network structure}

The model presented above imposes a structural constraint on the underlying network topology: in order for the evolution of the speakers' preferences to be well-defined, each of them must be part of, at least, one triangle. In fact, it has been recently shown that real social networks are characterized by an abundance of triangles, related to high values of the clustering coefficient \cite{Newman2003,Dorogovtsev2003,Newman2010,Foster2011,Colomer2013}. Thus, we consider networks with a large proportion of triangles \cite{Serrano2005,Newman2009,Bianconi2014}. In particular, we focus here on a socially inspired network generation algorithm proposed by Klimek and Thurner \cite{Klimek2013} and based on triadic closure, i.e., on the principle that individuals tend to make new acquaintances among friends of friends. Validated with data from a well-studied massive multiplayer online game \cite{Szell2010,Szell2010B,Szell2012,Klimek2016}, this network generation model involves three different mechanisms: random link formation, triadic closure \mbox{---link} formation between nodes with a common \mbox{neighbor---,} and node replacement \mbox{---removal} of a node with all its links and introduction of a new node with a certain number of \mbox{links---.}

\begin{figure}[ht]
    \centering
    \includegraphics[width=12.5cm, height=!]{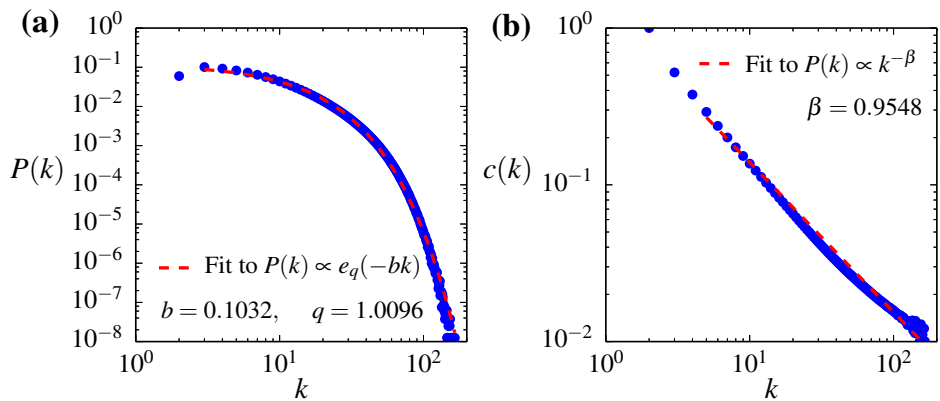}
    \caption{Panel \textbf{(a)}: Degree distribution and fit to a $q$-exponential function (for $k \geq 3$). Panel \textbf{(b)}: Average clustering coefficient as a function of the degree and fit to a power-law function (for $k \geq 5$). $10000$ realizations of the network generation algorithm were used.}
    \label{C6_f:degreeClusteringDistributions}
\end{figure}

Bearing in mind the structural constraint imposed by our model, and noticing that the node replacement mechanism might lead to some nodes losing all their triangles, we introduce a modification of the algorithm so as to avoid removing all the triangles from any node. Namely, when the removal of a node would lead to some of its neighbors losing all their triangles, these neighbors are arranged in triangles between themselves, or with randomly chosen nodes when necessary. Furthermore, the new node is introduced as a triangle by initially linking it with a random node and one of its neighbors. Finally, we use the same parameter values found by Klimek and Thurner \cite{Klimek2013} when calibrating their algorithm to the friendship network of the above-mentioned online game: a probability of triadic closure $c=0.58$ [being $(1-c)$ the probability of random link formation] and a probability of node replacement $r=0.12$. The degree distribution and the scaling of the average clustering coefficient as a function of the degree are shown in Fig.~\ref{C6_f:degreeClusteringDistributions} for the networks obtained in this manner. A fit of the degree distribution to a $q$-exponential function, \mbox{$P(k) \propto e_q(-bk)$} with \mbox{$e_q(x) = (1 + (1 - q)x)^{(1/(1-q))}$}, leads to a value of $q$ compatible with a purely exponential decay [$q=1.0096$, see panel \textbf{(a)}]. Regarding the average clustering coefficient as a function of the degree, a fit to a power-law decay leads to an exponent slightly smaller than one [$\beta=0.9548$, see panel \textbf{(b)}]. Comparing these results with those presented by Klimek and Thurner \cite{Klimek2013} (with fitting parameters $q=1.1162$ and $\beta=0.693$), we conclude that the described modification does not affect the general characteristics of the networks created, but it does have an effect on the specific values of the different scaling exponents.

\section{Transient dynamics and asymptotic configurations}
\label{C6_s:transient_dynamics_and_asymptotic_configurations}


By means of numerical simulations, we study the coupled dynamics of node and link states described above. Let us start by introducing three different measures characterizing the state of the system at any given time. Firstly, bearing in mind that we are interested here in the survival of languages regarding their actual use in the interactions between speakers, we can define an \emph{order parameter} $\rho$ in terms of link states. In particular, we define $\rho$ as the density of nodal interfaces \cite{FernandezGracia2012,Carro2014}, i.e., the fraction of pairs of connected links \mbox{---links} sharing a \mbox{node---} that are in different states,
\begin{equation}
    \rho = \frac{\displaystyle\sum_{i=1}^N k_i^{A} \, k_i^{B}}{\displaystyle\sum_{i=1}^N k_i (k_i - 1)/2} \, ,
    \label{C6_e:rho}
\end{equation}
where $k_i$ is the degree of node $i$, and $k_i^{A/B}$ is the number of $A/B$-links connected to node $i$. The order parameter $\rho$, by definition $\rho \in [0,1]$, is thus a measure of the local order in the system, becoming $\rho=0$ when all connected links share the same state and $\rho=1/2$ for a random distribution of link states. Note that, defined as such, the order parameter $\rho$ can also be understood as the usual density of active links \mbox{---fraction} of links connecting nodes with different \mbox{states---} in the line-graph of the original network \cite{Rooij1965,Chartrand1969,MankaKrason2010,Krawczyk2011,FernandezGracia2012,Carro2014}.


Secondly, we introduce the \emph{fraction of links in the minority language}, $m$, as an alternative, non-local measure characterizing the system in terms of link states,
\begin{equation}
    m =
    \begin{cases}
        \frac{\displaystyle\sum_{i} k_i^A}{\displaystyle\sum_{i} k_i}\, , & \text{if} \quad \displaystyle\sum_{i} k_i^A \leq \displaystyle\sum_{i} k_i^B\, ,\\[25pt]
        \frac{\displaystyle\sum_{i} k_i^B}{\displaystyle\sum_{i} k_i}\, , & \text{otherwise}\, .
    \end{cases}
    \label{C6_e:minorityLinkFrac}
\end{equation}
Note that the minority language is thereby defined as that which is less used in interactions between speakers \mbox{---fewer} links in the corresponding \mbox{state---,} regardless of the total number of those speakers. In this way, even if a majority of the population uses a certain language in some of their interactions, we will still consider it to be the minority language if only a minority of the total number of interactions actually take place in that language. By definition, $m \in [0, 1/2]$.


Finally, we can characterize the system in terms of node states by introducing the \emph{average preference of the speakers for the minority language}, $x^M$,
\begin{equation}
    x^M =
    \begin{cases}
        \displaystyle\frac{1}{N}\sum_{i} x_i\, , & \text{if} \quad \displaystyle\sum_{i} k_i^A \leq \displaystyle\sum_{i} k_i^B\, ,\\[20pt]
        \displaystyle\frac{1}{N}\sum_{i} (1 - x_i)\, , & \text{otherwise}\, ,
    \end{cases}
    \label{C6_e:avMinorityNodePref}
\end{equation}
where, as before, the minority language is identified according to the fraction of interactions in which it is used. By definition, $x^M \in [0,1]$.

\begin{figure}
    \centering
    \includegraphics[width=11cm, height=!]{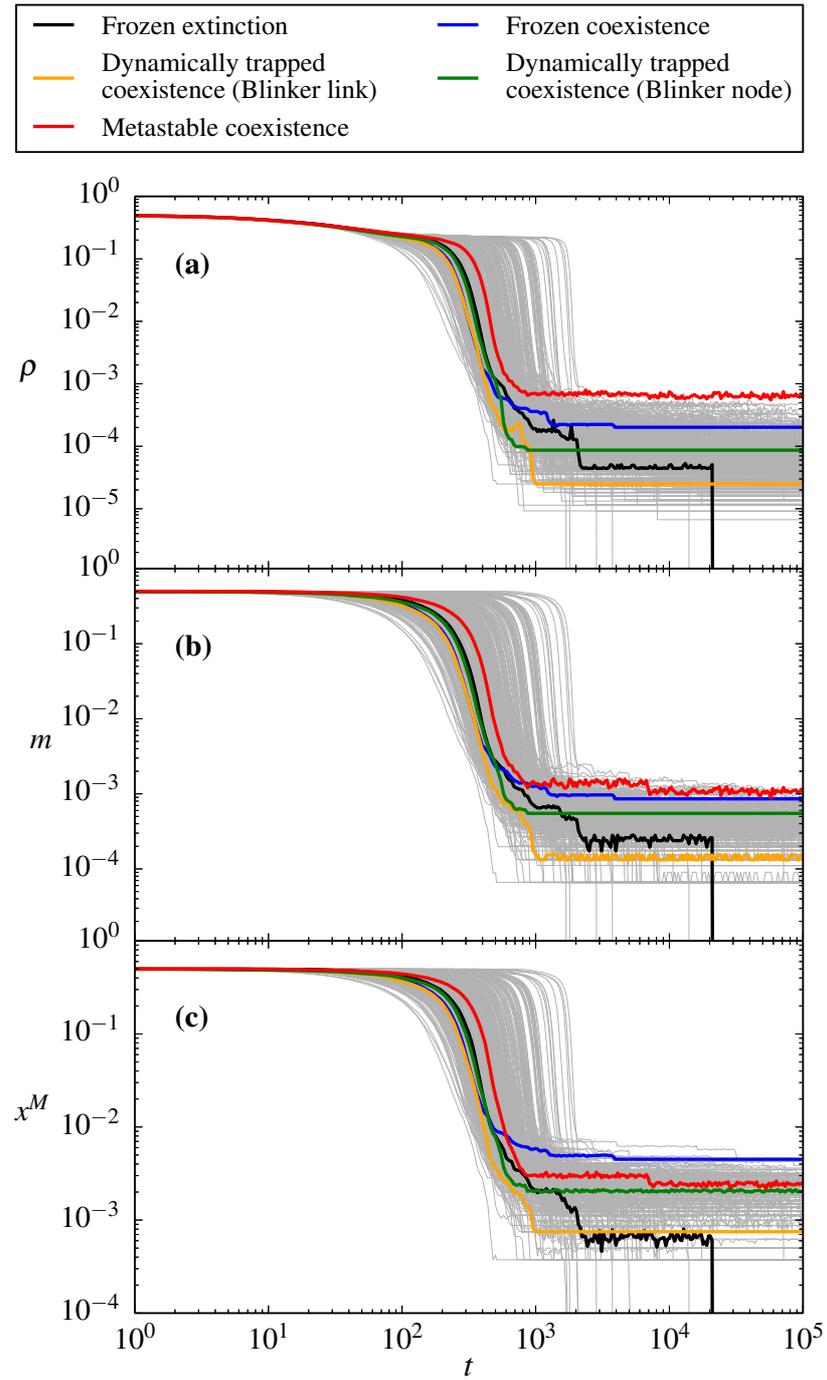}
    \caption{Time evolution of \textbf{(a)} the order parameter, \textbf{(b)} the fraction of links in the minority language, and \textbf{(c)} the average preference of the speakers for the minority language. $200$ individual realizations of the process are shown, among which $5$ realizations are highlighted as representative of the different types of possible trajectories. The system size used is $N = 8000$.}
    \label{C6_f:Time-Realizations}
\end{figure}
\begin{figure}[ht!]
    \centering
    \subfloat[Frozen coexistence\label{C6_f:AbsorbingCoexistence}]{
        \includegraphics[width=3.5cm, height=!]{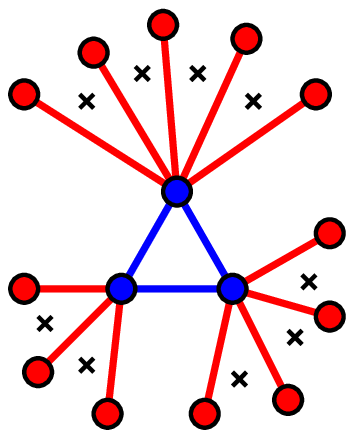}
    }
    \hskip1.75cm
    \subfloat[Dynamically trapped coexistence (blinker link)\label{C6_f:BlinkerLink}]{
        \includegraphics[width=3.5cm, height=!]{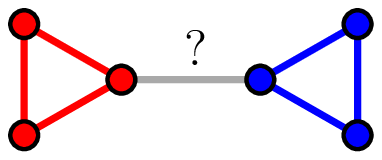}
    }
    \hskip1.75cm
    \subfloat[Dynamically trapped coexistence (blinker node)\label{C6_f:BlinkerNode}]{
        \includegraphics[width=3.5cm, height=!]{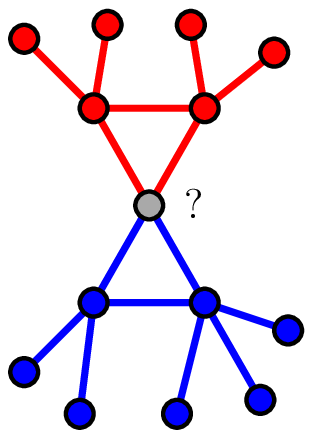}
    }
    \caption{Schematic illustration of the kind of structural motifs characterizing the different asymptotic configurations. The use of the two competing languages is represented, respectively, by solid red links and dashed blue links, while the preferences of the speakers are represented by node colors ranging from red to blue through white. Gray color is used to represent blinking or undecided situations. Crosses indicate the non-existence of a link.}
    \label{C6_f:AsymptoticConfigurations}
\end{figure}

The time evolution of these three measures is presented in Fig.~\ref{C6_f:Time-Realizations} for individual realizations of the model: the order parameter in panel~\emph{(a)}, the fraction of links in the minority language in panel~\emph{(b)}, and the average preference of the speakers for the minority language in panel~\emph{(c)}. All realizations start from a random initial distribution of states for both nodes and links, leading to all three measures starting from $1/2$, and they all experience a substantial ordering process in which one of the languages becomes predominant, leading to a large decrease of all three measures. Nevertheless, a variety of asymptotic behaviors can be observed. These behaviors are a direct consequence of the different types of asymptotic configurations reached by the system, which can be classified as:
\begin{enumerate}[label=(\emph{\roman*})]
    \item \emph{Frozen extinction states}: Absorbing configurations where one of the languages has completely disappeared, all links and nodes sharing the same state, and thus no further change of state is possible in the system. As a result, all the three introduced measures become zero (see black lines in Fig.~\ref{C6_f:Time-Realizations}).
    \item \emph{Frozen coexistence states}: Absorbing configurations where both language still exist but no further change of state is possible in the system. As a result, all our three measures remain constant with non-zero values (see blue lines in Fig.~\ref{C6_f:Time-Realizations}). Note that these situations of coexistence are characterized by one of the languages becoming a minority but persisting in the form of ``ghetto-like'' structures, defined as subsets of nodes such that all of them belong to triangles completely included in the subset. A schematic illustration of a simple ``ghetto-like'' motif composed of a single triangle can be found in Fig.~\ref{C6_f:AbsorbingCoexistence}.
    \item \emph{Dynamically trapped coexistence states}: Configurations where both languages still exist and the system is forever dynamic, but only a restricted (and usually small) number of changes of state are possible. In particular, only changes that do not modify the density of nodal interfaces are accessible. Bearing in mind that, by definition of the model, these changes are reversible, the system can move back and forth \emph{ad infinitum} \cite{Olejarz2011a,Olejarz2011b}. Depending on the kind of dynamical trap involved, we can identify three types of configurations:
    \begin{enumerate}[label={--}]
        \item Configurations based on \emph{Blinker links}: Both the order parameter and the average preference of the speakers for the minority language remain constant while the fraction of links in the minority language fluctuates around a certain value (see orange lines in Fig.~\ref{C6_f:Time-Realizations}). A schematic illustration of the most simple blinker link motif is presented in Fig.~\ref{C6_f:BlinkerLink}.
        \item Configurations based on \emph{Blinker nodes}: Both the order parameter and the fraction of links in the minority language remain constant while the average preference of the speakers for the minority language fluctuates around a certain value (see green lines in Fig.~\ref{C6_f:Time-Realizations}). A schematic illustration of a single blinker node motif can be observed in Fig.~\ref{C6_f:BlinkerNode}.
        \item Configurations based on both \emph{blinker links} and \emph{blinker nodes}: More or less complex combinations of the two previous types, leading to a constant order parameter and a fluctuating fraction of links in the minority language and average preference of the speakers for it.
    \end{enumerate}
    Note that dynamical traps can only appear at the interface between the frozen, ``ghetto-like'' structures described above and the rest of the network.
\end{enumerate}
Apart from these asymptotic configurations, we can also observe the presence of long-lived \emph{metastable coexistence states}. These non-trivial dynamical states are characterized by fluctuating, non-zero values of all the three introduced measures (see red lines in Fig.~\ref{C6_f:Time-Realizations}). The metastability of these states is based on a variation or weaker version of the ``ghetto-like'' structures described above, which would now consist of a subset of nodes such that a significantly large fraction of the triangles they belong to are completely included in the subset, i.e., they have a significantly larger number of triangles towards the inside of the subset than towards the outside. Due to finite-size fluctuations, however, the system always ends up falling to one of the previously described asymptotic states.

\begin{figure}[ht!]
    \centering
    \includegraphics[width=8.75cm, height=!]{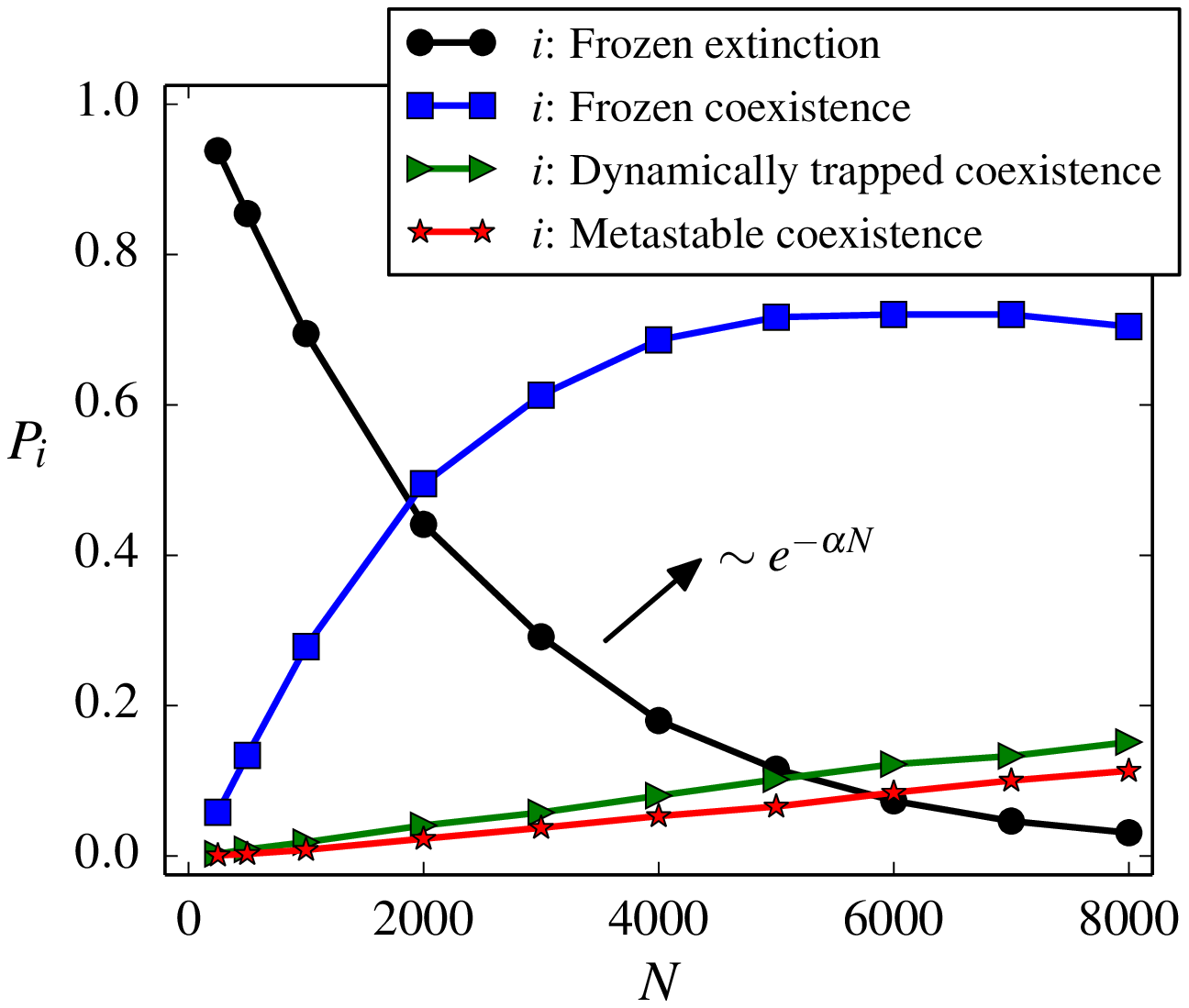}
    \caption{Scaling with system size of the fraction of realizations having reached each of the possible asymptotic configurations at time $t=10^5$, as well as the fraction of those still in a metastable state. A total of $10000$ realizations were used, with different networks and different initial conditions.}
    \label{C6_f:SizeScaling-FinalFractions}
\end{figure}

Once the different types of asymptotic states of the system have been presented, let us now focus on their relative likelihood. In particular, we show in Fig.~\ref{C6_f:SizeScaling-FinalFractions} the fraction of realizations having reached each of the possible asymptotic configurations before the end of the studied time period ($t = 10^5$), as well as the fraction of those still in a metastable state, for different system sizes. While the (frozen) extinction of the minority language is the most likely outcome for small systems ($N < 2000$), its probability decreases exponentially with system size, thus becoming negligible for large enough systems. Frozen coexistence is clearly predominant for large system sizes inside the studied range (\mbox{$2000 < N \leq 8000$}). However, given the linear growth observed in the fraction of dynamically trapped coexistence configurations, the numerical results presented in this figure for limited system sizes are inconclusive regarding the prevalence of frozen or dynamically trapped coexistence in the infinite size limit. Regarding the metastable coexistence states, it should be noted that they are not asymptotic states, and thus the system will eventually end up falling to any of the other frozen or dynamically trapped configurations. The fact that the fraction of metastable realizations at a given time grows linearly with the system size, suggests that the time scale in which the system is able to leave those metastable states also grows linearly with $N$, a point that will be discussed further in the next section.

\section{Time scales of extinction and metastable coexistence}
\label{C6_s:time_scales_of_extinction_and_metastable_coexistence}


Due to the diversity of possible asymptotic configurations described in the previous section, different time scales can be defined to characterize the dynamics of the system. In particular, we focus here on two time scales: the characteristic time of extinction of one of the languages and the characteristic duration or survival time of the metastable states. While in the first case we focus on realizations reaching the frozen extinction state over the time period under study, in the second case we consider all realizations leaving the metastable coexistence state over that time period, regardless of the particular asymptotic state they reach.

\begin{figure}[ht!]
    \centering
    \includegraphics[width=8.8cm, height=!]{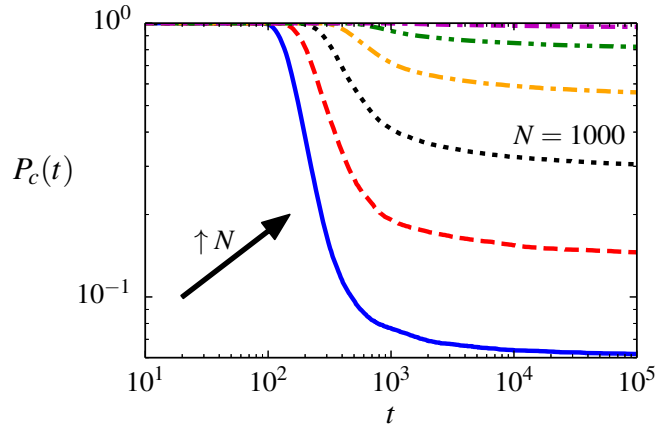}
    \caption{Time evolution of the coexistence probability (fraction of realizations not having reached the frozen extinction configuration by time $t$) for different system sizes, namely $N=250$, $500$, $1000$, $2000$, $4000$, and $8000$. A total of $10000$ realizations were used, with different networks and different initial conditions.}
    \label{C6_f:Time-coexistenceProb}
\end{figure}

Let us start by considering the time evolution of the probability of coexistence of both languages $P_c(t)$, i.e., the fraction of realizations not having reached a frozen extinction configuration by time $t$, depicted in Fig.~\ref{C6_f:Time-coexistenceProb} for different system sizes. Coherent with the results presented above in Fig.~\ref{C6_f:SizeScaling-FinalFractions}, the coexistence probability becomes closer and closer to one, for any time, as the system size becomes larger and larger. For small systems, on the contrary, the probability of both languages coexisting shows a large decrease around a certain characteristic time, which grows with system size, before asymptotically reaching a plateau. Note, nevertheless, that this plateau is not reached as long as there are metastable realizations able to reach the frozen extinction configuration.

\begin{figure}[ht!]
    \centering
    \includegraphics[width=8.9cm, height=!]{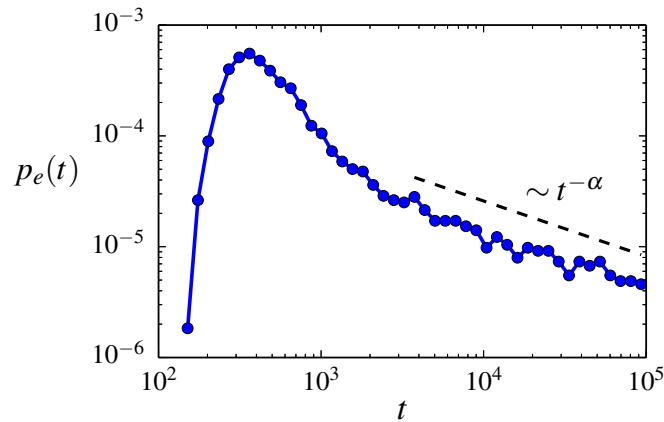}
    \caption{Distribution of extinction times for realizations reaching the frozen extinction configuration over the interval $t \in [0,10^5]$ ($69.5\%$ of the $10000$ realizations studied) for a system size $N=1000$. A power-law decay with exponent $\alpha = 0.5$ is shown as a guide to the eye.}
    \label{C6_f:OrderingTime-Dist}
\end{figure}

As we can observe in Fig.~\ref{C6_f:Time-coexistenceProb}, most extinction events take place around a certain characteristic time. For instance, for the system size $N=1000$, $90\%$ of all extinction events observed in the interval $t \in [0,10^5]$ take place between $t=200$ and $t=2000$. However, for a non-negligible fraction of realizations the extinction of one of the languages happens at significantly longer times, and thus the coexistence probability keeps on slowly decreasing instead of quickly reaching a plateau. In order to further analyze this behavior, we present in Fig.~\ref{C6_f:OrderingTime-Dist} the probability distribution of extinction times $p_e(t)$ for the system size $N=1000$, where, according to the results presented in Fig.~\ref{C6_f:SizeScaling-FinalFractions}, extinction is predominant. Note that this distribution is related to the coexistence probability by
\begin{equation}
    P_c(t) = 1 - \int_0^{t} p_e(t')dt' \, .
\end{equation}
In this way, we see that extinction times are broadly distributed and that the decay of their probability for long times seems to be compatible with a power law $p_e(t) \sim t^{-\alpha}$ with exponent $\alpha \sim 0.5$. Being the exponent smaller than one, the mean of the distribution diverges, and thus there is no well-defined characteristic time scale for the extinction events. As a consequence, even if the extinction of one of the languages is predominant for small system sizes, there are, at all time scales, realizations where both languages are still coexisting.

\begin{figure}[ht!]
    \centering
    \includegraphics[width=8.9cm, height=!]{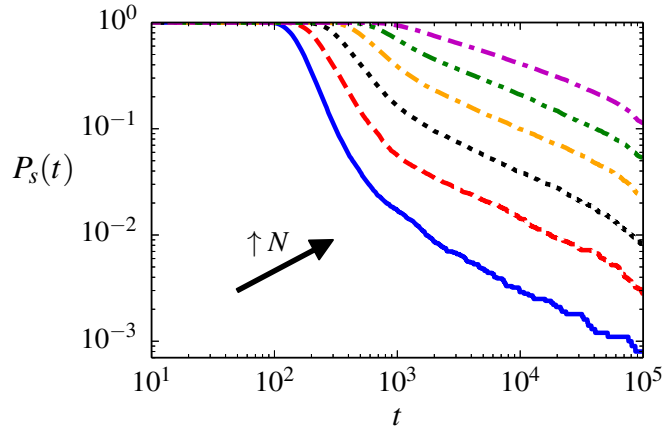}
    \caption{Time evolution of the survival probability of the metastable states (fraction of realizations not having reached any frozen or dynamically trapped state by time $t$) on a log--log scale and for different system sizes, namely $N=250$, $500$, $1000$, $2000$, $4000$, and $8000$. A total of $10000$ realizations were used, with different networks and different initial conditions.}
    \label{C6_f:Time-survivalProb-LogLog}
\end{figure}

A further characterization of the behavior of the model is given by the time scale at which the system is able to escape from the metastable coexistence states, i.e., the characteristic survival time of these non-trivial dynamical states before the dynamics of the system becomes locked in any frozen or dynamically trapped configuration. In order to study this, let us first introduce the survival probability of the metastable states $P_s(t)$, defined as the fraction of realizations not having reached any frozen or dynamically trapped state by time $t$. Our results for this probability are presented in Fig.~\ref{C6_f:Time-survivalProb-LogLog}, on a log--log scale, for different system sizes. Comparing Figs.~\ref{C6_f:Time-coexistenceProb} and~\ref{C6_f:Time-survivalProb-LogLog} we can observe that, similarly to the coexistence probability, surviving realizations are more and more likely, for any point in time, for larger and larger systems. On the contrary, the survival probability of the metastable states does not asymptotically approach any plateau, as it was the case for the coexistence probability. This is coherent with the fact that, by definition, all metastable realizations eventually end up being frozen or dynamically trapped.

\begin{figure}[ht!]
    \centering
    \includegraphics[width=8.9cm, height=!]{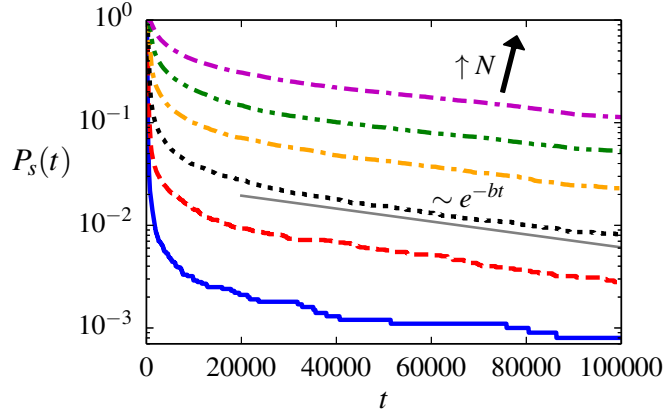}
    \caption{Time evolution of the survival probability of the metastable states (fraction of realizations not having reached any frozen or dynamically trapped state by time $t$) on a semilogarithmic scale and for different system sizes, namely $N=250$, $500$, $1000$, $2000$, $4000$, and $8000$. An exponential decay with the slope obtained by fitting the data for $N=1000$ is also shown as a guide to the eye (thin solid gray line). A total of $10000$ realizations were used, with different networks and different initial conditions.}
    \label{C6_f:Time-survivalProb-LogLin}
\end{figure}

Even if the survival probability of the metastable states appears to be fat-tailed in the log--log scale of Fig.~\ref{C6_f:Time-survivalProb-LogLog}, a closer look at the same results presented on a semilogarithmic scale in Fig.~\ref{C6_f:Time-survivalProb-LogLin} shows that any fat-tailed behavior is interrupted by an exponential decay occurring after a long cutoff time. This final exponential decay allows for both the mean and the fluctuations of the distribution of survival times of the metastable states to be well-defined, and thus the mean can play the role of a characteristic duration or survival time of these metastable states $\tau_s$ before the system reaches a frozen or dynamically trapped configuration.

The characteristic survival time of the metastable states $\tau_s$ can be directly computed from their survival probability $P_s(t)$ as
\begin{equation}
    \tau_s = \int_0^{\infty} P_s(t) dt \, .
    \label{C6_e:tau_s_definition}
\end{equation}
However, given that a non-negligible number of realizations in our sample stay in a metastable state for the whole period under study \mbox{---particularly} for large system \mbox{sizes---,} we cannot simply discard the queue of the distribution and numerically compute the mean using only the observed survival times. On the contrary, we need to take the queue of the distribution into account, which we can do by fitting the final exponential decay uncovered above in Fig.~\ref{C6_f:Time-survivalProb-LogLin}. In particular, if we assume that the survival probability of the metastable states takes the functional form
\begin{equation}
    P_s(t) = P_s(t^*) e^{-b(t-t^*)} \quad \textrm{for } t \geq t^* \, ,
\end{equation}
from a certain cutoff time $t^*$, where $b$ and $t^*$ are fitting parameters, then we can divide the integral in Eq.~\ref{C6_e:tau_s_definition} into two terms,
\begin{equation}
    \tau_s = \int_0^{t^*} P_s(t) dt + \int_{t^*}^{\infty} P_s(t) dt = \int_0^{t^*} P_s(t) dt + \int_{t^*}^{\infty} P_s(t^*) e^{-b(t-t^*)} dt \, .
\end{equation}
Finally, performing the integral in the last term, we find an expression for the characteristic survival time of the metastable states as a sum of two contributions,
\begin{equation}
    \tau_s = \int_0^{t^*} P_s(t) dt + \frac{P_s(t^*)}{b} \, ,
\end{equation}
the first of which can be numerically computed as the average of the survival times of the realizations reaching a frozen or dynamically trapped state before $t^*$. Regarding the second contribution, it can be computed by an exponential fit to the results presented in Fig.~\ref{C6_f:Time-survivalProb-LogLin} for $t \geq t^*$.

\begin{figure}[ht!]
    \centering
    \includegraphics[width=8.7cm, height=!]{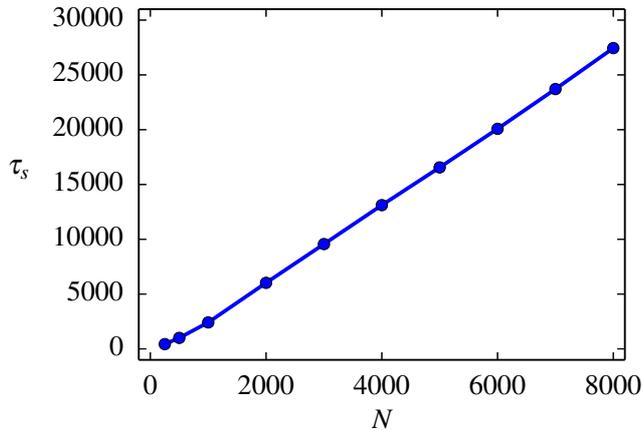}
    \caption{Scaling with system size of the characteristic survival time of the metastable states. A total of $10000$ realizations were used, with different networks and different initial conditions.}
    \label{C6_f:SizeScaling-SurvivalTime}
\end{figure}

Results for the scaling with system size of the characteristic survival time of the metastable states are presented in Fig.~\ref{C6_f:SizeScaling-SurvivalTime}, showing a linear relationship between both quantities. Thus, for increasing system sizes, realizations survive for longer and longer times in a metastable state before falling to a frozen or dynamically trapped configuration. Moreover, in the infinite size limit, the system is unable to escape from the metastable states in any finite time.

\section{Use of the minority language}
\label{C6_s:Use_of_the_minority_language}

\begin{figure}[ht!]
    \centering
    \includegraphics[width=10.5cm, height=!]{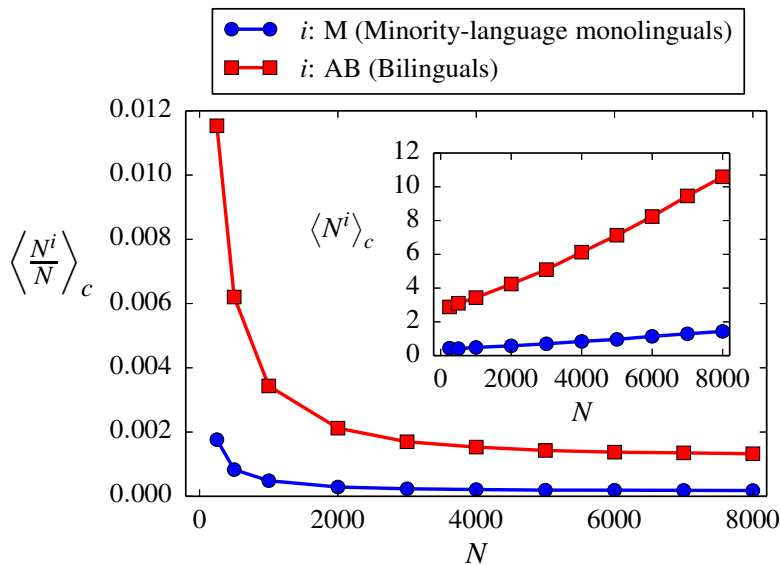}
    \caption{Scaling with system size of the fractions of minority-language monolingual and bilingual speakers at the last point of the time period under study, $t=10^5$, averaged over coexisting realizations. Inset: Scaling with system size of the absolute number of minority-language monolingual and bilingual speakers at the last point of the time period under study, $t=10^5$, averaged over coexisting realizations.}
    \label{C6_f:SizeScaling-Monolinguals-Bilinguals}
\end{figure}

Once we have identified the different types of configurations associated with the coexistence of both languages, and studied their probabilities and typical time scales, let us now turn our attention to the extent of this coexistence. In particular, bearing in mind that the situations of coexistence are characterized by one of the languages becoming a clear minority (see Section~\ref{C6_s:transient_dynamics_and_asymptotic_configurations}), we focus here on two measures quantifying the use of this minority language: the number of speakers who use only this language (minority language monolingual speakers, $N^M$) and the number of those who use both languages (bilingual speakers, $N^{AB}$). In order to consider only very long-lived metastable states, apart from frozen and dynamically trapped coexistence configurations, we focus on the last point of the time period under study, $t=10^5$, and we average only over realizations where both languages are still coexisting (which we note by $\langle \cdot \rangle_c$). Results for the dependence of these two quantities on system size are presented in Fig.~\ref{C6_f:SizeScaling-Monolinguals-Bilinguals}, measured as fractions of the total number of speakers in the main plot and as absolute numbers in the inset.

A first observation is that the state of coexistence is predominantly sustained by bilingual speakers, both their fraction and their absolute number being significantly larger that those corresponding to monolingual speakers of the minority language for any system size. Secondly, while both fractions of minority-language speakers are shown to be decreasing functions of the system size for small systems, they appear to be reaching a plateau for large systems. On the one hand, bearing in mind that averages are computed over coexisting realizations \mbox{---rare} for small systems, predominant for large \mbox{ones---,} this suggests that there is a minimum size of the structures sustaining the use of the minority language. In this way, the smaller the size of the system, the less likely these minimal structures are to appear, but the larger the fraction of the system they represent whenever they actually appear. On the other hand, the asymptotic tendency of both fractions towards a plateau suggests a linear growth with system size of the absolute number of both monolingual and bilingual speakers of the minority language for large systems, which is confirmed in the inset. Finally, the substantially faster growth of the absolute number of bilingual speakers with system size, as compared to minority-language monolinguals, underlines again the importance of bilinguals in sustaining the use of the minority language: bilingualism becomes more and more prevalent among speakers of the minority language for growing systems.


\section{Comparison with the AB-model}
\label{C6_s:Comparison_with_the_AB-model}


Given the non-standard topology used for our numerical simulations, imposed by the structural constraints of the model \mbox{---namely,} an abundance of \mbox{triangles---,} we present here, for comparison, numerical results for the \mbox{AB-model} in the same networks. Let us first briefly recall the main features of this model, in which language use is considered to be a state of the agents. As outlined in Section~\ref{C6_s:introduction}, the \mbox{AB-model} was proposed by Castell\'{o} \emph{et al} \cite{Castello2006} based on the works of Wang and Minett \cite{Wang2005}, and it develops a modification of the original, binary-state Abrams-Strogatz model to account for the case of two non-excluding options by introducing a third, intermediate state. Thus, agents can be in one of the following states: $A$ (monolingual speaker of language $A$), $B$ (monolingual speaker of language $B$), or $AB$ (bilingual speaker). Starting from a random initial distribution of states, an agent is randomly chosen at each iteration and its state is updated according to the following probabilities,
\begin{equation}
\begin{aligned}
    &p_{A \to AB} = \frac{1}{2} \sigma_B \, , &\qquad &p_{B \to AB} = \frac{1}{2} \sigma_A \, ,\\[5pt]
    &p_{AB \to B} = \frac{1}{2} (1 - \sigma_A) \, , &\qquad &p_{AB \to A} = \frac{1}{2} (1 - \sigma_B) \, ,
\end{aligned}
\end{equation}
where $\sigma_A$, $\sigma_B$ and $\sigma_{AB}$ are, respectively, the fractions of neighbors of the chosen agent in state $A$, $B$ and $AB$ (note that \mbox{$\sigma_A + \sigma_B + \sigma_{AB} = 1$}). That is, monolingual speakers of $A$ ($B$) become bilinguals with a probability proportional to the local fraction of monolingual speakers of $B$ ($A$), while bilinguals become monolingual speakers of $A$ ($B$) with a probability proportional to the local fraction of speakers of $A$ ($B$), which includes both monolingual and bilingual speakers.

\begin{figure}[ht!]
    \centering
    \includegraphics[width=8.6cm, height=!]{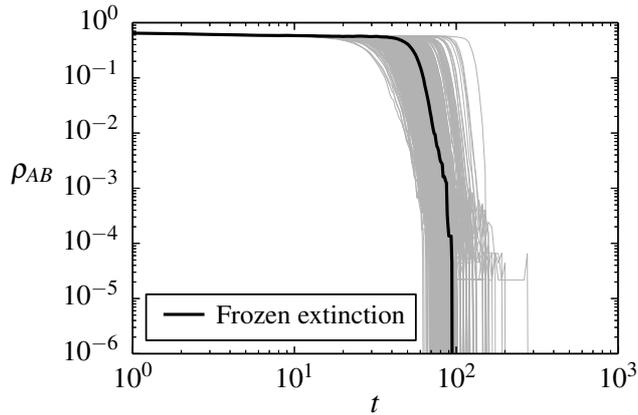}
    \caption{Time evolution of the order parameter (interface density) for the AB-model. $200$ individual realizations of the process are shown. The system size used is $N = 8000$.}
    \label{C6_f:AB_Time-InterfaceDens}
\end{figure}

Frozen coexistence configurations and dynamically trapped states are not possible in the AB-model, which has, by definition, a single absorbing state: the extinction of one of the languages. Therefore, the order parameter $\rho_{AB}$, defined now as the density of link interfaces \mbox{---fraction} of links connecting nodes with different \mbox{states---,} is enough to characterize the time evolution of individual realizations, some of which are shown in Fig.~\ref{C6_f:AB_Time-InterfaceDens}. While this parameter $\rho_{AB}$ is different from the order parameter $\rho$ used above to characterize our model, both of them are measures of the local order of the system. Note that, due to the existence of three different states, all realization start from $\rho_{AB}=2/3$, corresponding to a random initial distribution of states. Similarly to our model, all the realizations go through a substantial ordering process in which one of the languages becomes predominant. In contrast to our model, however, this ordering process takes place around an order of magnitude before ($t \sim 10^2$ as opposed to $t \sim 10^3$) and it very quickly leads to the complete extinction of one of the languages, all nodes sharing the same state, whether $A$ or $B$ monolingual. Furthermore, only very few realizations are observed to last noticeably longer than the rest of them ($t \sim 600$), suggesting that there are no long-lived metastable coexistence states (compare with Fig.~\ref{C6_f:Time-Realizations}).

\begin{figure}[ht!]
    \centering
    \includegraphics[width=8.6cm, height=!]{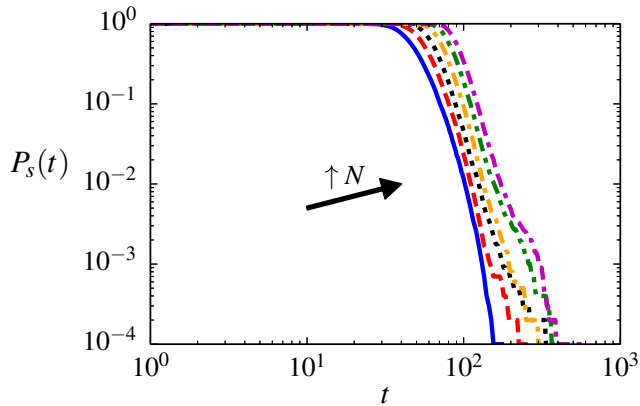}
    \caption{Time evolution of the survival probability (fraction of realizations not having reached a frozen state by time $t$) for the AB-model and for different system sizes, namely $N=250$, $500$, $1000$, $2000$, $4000$, and $8000$. A total of $10000$ realizations were used, with different networks and different initial conditions.}
    \label{C6_f:AB_Time-survivalProb}
\end{figure}

Given that the only asymptotic state of the AB-model is the frozen extinction of one of the languages, the survival of a non-trivial dynamics \mbox{---not} having reached any frozen or dynamically trapped \mbox{state---} and the coexistence of both languages \mbox{---not} having reached the frozen extinction \mbox{state---} are equivalent, and so are their respective probabilities, $P_s$ and $P_c$. Results for the time evolution of the survival (or coexistence) probability $P_s(t)$ are presented in Fig.~\ref{C6_f:AB_Time-survivalProb} for different system sizes. As we can observe, after a very short transient (lasting until $t \sim 30$), the likelihood of an active state where both languages coexist quickly falls to zero, with no fat-tailed behavior. Furthermore, this decrease seems to be almost independent of system size. Both features are in agreement with the results reported for the AB-model in random networks without communities \cite{Castello2007,Toivonen2009}. They are, however, in sharp contrast with the results corresponding to our model, presented above in Figs.~\ref{C6_f:Time-coexistenceProb} and~\ref{C6_f:Time-survivalProb-LogLog}.

\begin{figure}[ht!]
    \centering
    \includegraphics[width=10.2cm, height=!]{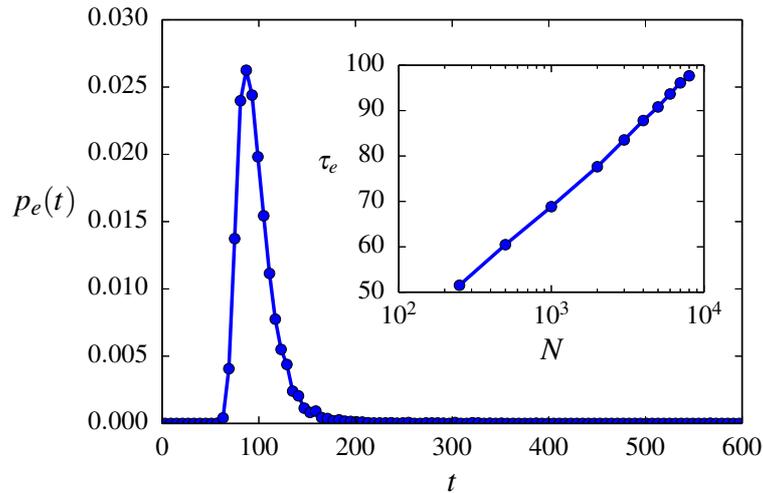}
    \caption{Distribution of extinction times for the AB-model and for a system size $N=8000$. Inset: Scaling with system size of the characteristic extinction time for the AB-model. A total of $10000$ realizations were used, with different networks and different initial conditions.}
    \label{C6_f:AB_OrderingTime-Dist_SizeScaling}
\end{figure}

The probability distribution of extinction (or survival) times $p_e(t)$ is shown in Fig.~\ref{C6_f:AB_OrderingTime-Dist_SizeScaling} for a system size $N=8000$. As opposed to the results presented in Fig.~\ref{C6_f:OrderingTime-Dist} for our model, the extinction times of the AB-model are very closely distributed around the peak, i.e., almost no realization is found to last significantly longer than the rest of them \mbox{---suggesting} the absence of long-lived metastable \mbox{states---.} Therefore, the mean of the distribution is well-defined and it can be used as a characteristic extinction time scale. In contrast with the method used to analyze our model, where a non-negligible number of realizations survived in a non-trivial dynamical state for the whole period of time under study (see Section~\ref{C6_s:time_scales_of_extinction_and_metastable_coexistence}), we can here numerically compute the mean of the distribution from our sample of realizations, given that all their survival times are smaller that the studied time period. The dependence of this characteristic extinction time $\tau_e$ on system size is also depicted in Fig.~\ref{C6_f:AB_OrderingTime-Dist_SizeScaling} as an inset. In particular, $\tau_e$ is found to be a logarithmic function of the system size, to be compared with the linear relationship found for our model and shown in Fig.~\ref{C6_f:SizeScaling-SurvivalTime}. This result is coherent with the previous observation regarding the small influence of system size on the survival probability.

\section{Summary and conclusions}
\label{C6_s:Conclusions}


We introduced a prototype model of coupled evolution of node and link states. In particular, we implemented a language competition model where both the use and the preference for a given language are included, with different but coupled dynamics. We proposed to consider the use of a language as a property of the interactions between speakers \mbox{---a} link \mbox{state---} and the preference or attitude of the speakers towards it as a property of the speakers themselves \mbox{---a} node \mbox{state---.} Furthermore, we focused on the case of two socially equivalent languages and performed numerical simulations on socially inspired network topologies based on a mechanism of triadic closure. As opposed to most of the previously proposed models, where the extinction of one of the languages is an inevitable outcome of the dynamics, we found a broad range of possible asymptotic configurations, which can be classified as: frozen extinction states, frozen coexistence states, and dynamically trapped coexistence states. Furthermore, metastable coexistence states with non-trivial dynamics were found to be abundant and with very long survival times. By means of a system size scaling, we showed that the probability of extinction of one of the languages decreases exponentially with system size, therefore becoming negligible for large enough systems. Moreover, we showed that, even for small systems, extinction times are so broadly distributed that coexisting realizations can be found at all time scales. Regarding the metastable coexistence states, we showed that their characteristic survival time before the system reaches any frozen or dynamically trapped configuration scales linearly with system size. Thus, in the infinite size limit, all realizations will be found to be in a non-trivial dynamical coexistence state for any finite time. Finally, we showed that bilingualism becomes more prevalent among speakers of the minority language the larger the size of the system.


The dynamics of the system being characterized by the fast emergence of a predominant language, we found that, as the use of the minority language decreases, it becomes increasingly confined to the more intimate social spheres or group interactions \mbox{---triangular} \mbox{relationships---.} In particular, the situations of coexistence (frozen, dynamically trapped, and metastable) were found to be based on the existence of ``ghetto-like'' structures, where predominantly bilingual speakers use the minority language for the interactions among themselves \mbox{---mostly} \mbox{triangular---} while they switch to the majority language for communications with the rest of the population \mbox{---mostly} \mbox{non-triangular---.} In this way, bilingual speakers with a strong preference for the minority language, and using it for their close group interactions, are found to play an essential role in its survival. Our results highlight the importance of the network topology for determining the possibility of coexistence of two competing languages. However, as opposed to previous studies, we find that group interactions \mbox{---in} the form of \mbox{triangles---} can play a more relevant role than simple one to one interactions.


A natural extension of the model presented here would be to consider a coevolving or dynamic topology \cite{Carro2014}, allowing to capture phenomena such as births, deaths, migration, and the evolution of social ties. The robustness of the above-described ``ghetto-like'' structures could, in this way, be studied for different rewiring rules. A different way of exploring this robustness would be to consider the effect of a certain type noise affecting the decisions of the speakers \cite{Diakonova2015,Carro2016}.


The ideas and methods presented here to study language competition processes can also be useful in different contexts. Indeed, the idea of a coevolution of node and link states is very general and could be applied whenever there is a relevant property associated to the interactions between agents or nodes and this property is characterized by a dynamics of its own, which is not completely determined by the states of the agents and their particular dynamics. Examples range from friendship-enmity relationships and trust to the coupled dynamics of trade and economic growth \cite{Garlaschelli2007}. Finally, the importance of triangular structures both in the definition and in the results of the model presented here calls for a generalization of the concept of network beyond the traditional dyadic interactions, in order to take into account also group interactions of higher order (triadic, tetradic, etc) \cite{Perc2013}.


\begin{acknowledgments}

We acknowledge financial support by FEDER (EU) and MINECO (Spain) under Grant ESOTECOS (FIS2015-63628-C2-2-R). AC acknowledges support by the FPU program of MECD.

\end{acknowledgments}



\bibliographystyle{iopart-num}
\bibliography{Bibliography.bib}

\end{document}